\documentclass[aps,preprint]{revtex4}
\usepackage{amssymb}
\usepackage{amsmath}
\usepackage{graphicx}
\usepackage{epstopdf}
\begin{document}

\title{Black Holes in  (Quartic) Quasitopological Gravity}
\author{M. H. Dehghani$^{1,2}$ \footnote{email address: mhd@shirazu.ac.ir},
A. Bazrafshan $^{3,1}$, R. B. Mann$^{4}$ \footnote{email address:
rbmann@sciborg.uwaterloo.ca}, M. R. Mehdizadeh$^{1}$, M. Ghanaatian$^{5}$ and M. H. Vahidinia$^{1}$}
\affiliation{1. Physics Department and Biruni Observatory, College of Sciences, Shiraz University, Shiraz 71454, Iran\\
2. Research Institute for Astrophysics and Astronomy of Maragha (RIAAM), Maragha, Iran\\
3. Department of Physics, Jahrom University, P.O. Box 74135-111 Jahrom, Iran\\
4. Department of Physics \& Astronomy, University of Waterloo, 200 University Avenue West, Waterloo, Ontario, Canada, N2L 3G1\\
5. Department of Physics, Payame Noor University, P.O. Box 19395-3697 Tehran, Iran}
\begin{abstract}
We construct quartic quasitopological gravity, a theory of gravity containing terms quartic in the curvature that yields second order differential equations in the spherically symmetric case.   Up to a term proportional to the
quartic term in Lovelock gravity we find a unique solution for this quartic case, valid in any dimensionality larger than 4 except 8.  This case is the highest degree of curvature coupling for which explicit black hole solutions can be constructed, and we obtain and analyze the various black hole solutions that emerge from the field equations in $(n+1)$ dimensions.  We discuss the thermodynamics of these black holes and compute their entropy as a function of the horizon radius.   We then make some general remarks about $K$-th order quasitopological gravity, and point out that the basic structure of the solutions will be the same in any dimensionality for general $K$ apart from particular cases.
\end{abstract}
\maketitle

\section{Introduction}

The gauge/gravity idea is that gravitational dynamics in a given
dimensionality can be mapped onto some other (nongravitational) field theory
of a lower dimensionality. The duality between central charges and couplings
on the nongravitational side and the parameters on the gravitational side
has been explored primarily through the trace anomaly \cite{Sken}. However
Einstein gravity does not have enough free parameters to account for the
ratios between central charges and therefore is only dual to those conformal
field theories for which all the central charges are equal. To broaden the
universality class of dual field theories which one can study with
holography, one must extend to more general theories of gravity which
contain more free parameters such as Lovelock theory \cite{Lov} or
quasitopological gravity \cite{OlivaRay,Myers1}. These additional central
charges have recently been investigated holographically \cite{Myers2,Myers3}.

Another point which is interesting in gauge/gravity duality is that
the dual CFT should respect causality. This creates a constraint on the coupling constants of
the gravity theory. In this analysis, one considers graviton fluctuations that probe the bulk geometry
in the presence of a black hole. In general the dual CFT
plasma may support superluminal signals, and so the gravitational couplings must be constrained so as
to avoid the appearance of such superluminal modes. For Lovelock gravity,
while  causality constraints precisely match those
arising from requiring positive energy fluxes \cite{Lg2, Lg3},
 it has been shown that this matching does not appear in general, specifically,
for cases where the gravitational equations of motion are not second order \cite{Hof}.
However for cubic quasitoplogical gravity there are three constraints that arise from requiring positive energy fluxes,
which determine the three coupling constants. No evidence for
causality violation was found once the curvature-cubed coupling was chosen consistent with these constraints \cite{Myers2}.

Motivated by the success of holographic studies of second \cite{Lg2}, and third-order Lovelock gravity \cite{Lg3,ThermoLov} and curvature-cubed, or cubic
quasitopological gravity \cite{Myers2,Myers3,Wilson}, we consider here
adding a quartic curvature term with a new coupling constant on the gravity
side, affording exploration of a larger space of field theories.
In quasitopological gravity, the linearized equations in a black hole background
are fourth-order
 in derivatives and so one does not expect  causality constraints
to match those arising from requiring positive energy fluxes.  In view of the results for
the cubic case \cite{Myers2}, the simplest nontrivial
case to consider is the quartic case; with this new coupling constant, we have four coupling constants and therefore
the constraints arising from  causality may   not match the three constraints arising
from requiring positive energy fluxes.  The first step in such an investigation is to
construct the quartic theory and analyze its basic properties.  We shall consider
the more detailed considerations of positivity of energy and causality in future work.

Progress with Lovelock gravity and cubic quasitopological gravity relies on the fact
that even though this is a higher curvature theory of gravity, the
holographic calculations in this model are still under control, at least in
spherically symmetric settings. This control in turn is based on the two
facts that the equations of motion are only second-order in derivatives
(again, for spherical symmetry) and that exact black hole solutions have
been constructed. Hence we want to introduce a quartic curvature topological
gravity for which the equation of motions are still second-order and exact
black hole solutions can be constructed.  In this context,
 fourth-order Lovelock and quasitoplogical  gravity is the largest order for which the
 field equations can be solved analytically.  Even in Lovelock gravity, this largest analytic solution
has not yet been considered. Here we carry out the first steps along these lines, studying the exact spherically
symmetric  solutions and their properties.

To obtain the quartic case,
a natural generalization would be to add interaction terms quartic in curvature
via fourth-order Lovelock gravity. However, because of the topological
origin of the Lovelock terms the quartic interaction term of Lovelock
gravity only contributes to the equations of motion when the bulk dimension
is nine or greater. In the context of  gauge/gravity duality, this means
that such a term will be effective in expanding the class of dual field
theory in eight or more dimensions. Our key result in this paper is to
construct a new gravitational action with quartic curvature interactions
(quartic quasitopological gravity) valid in lower dimensions, thereby providing
 a useful toy model to
study a broader class of four (and higher) dimensional CFTs, involving four
independent parameters.

Here we explicitly construct quartic quasitopological gravity in any
dimensionality except 8, beginning with the five dimensional case. Although
an action quartic in curvature terms has been previously constructed \cite
{OlivaRay} (and from which was proved a generalized Birkhoff theorem, namely
that constant
spherical/planar/hyperbolic transverse curvature implies staticity \cite
{Oliva:2011xu}), the field equations in the spherically symmetric case
vanish  in less than seven spacetime dimensions. In contrast to this, the
quartic topological action we construct yields nontrivial second-order field
equations in all spacetime dimensionalities but 8. Indeed, our quartic
curvature action differs from that of ref. \cite{OlivaRay} in terms of its
various coefficients, and insofar as it yields nontrivial field equations in
5 dimensions and higher.

We also present and discuss exact black hole solutions
of this new theory for various asymptotic boundary conditions.  These
solutions share a number of features in common with solutions from higher-order Lovelock
theories in greater dimensions.  For example, in the spherically symmetric cases we consider, the
field equations for the metric function in our quartic theory in 5 dimensions are formally the same as for
fourth-order Lovelock theory in 9 dimensions, differing only by the power of $r$ present in the resultant quartic
equation.
We furthermore consider the thermodynamic behavior of these objects for general
dimensionality. We leave a detailed study of the properties of the dual
class of field theories for future investigation.

\section{Quartic Topological Action in Five Dimensions\label{QT4}}

Motivated by considerations of the AdS/CFT correspondence, we want to
consider a curvature-quartic theory of gravity in five dimensions. We are
interested in a gravity theory which produces second-order equation of
motion and can have exact solutions. A natural candidate that has these
properties is the fourth-order Lovelock gravity with action
\begin{equation}
I_{\mathrm{G}}=\frac{1}{16\pi }\int d^{n+1}x\sqrt{-g}[-2\Lambda +\mathcal{L}%
_{1}+\alpha _{2}\mathcal{L}_{2}+\alpha _{3}\mathcal{L}_{3}+\alpha _{4}%
\mathcal{L}_{4}],  \label{Act1}
\end{equation}
where $\Lambda =-n(n-1)/2l^{2}$ is the cosmological constant for AdS
solutions, and the $\alpha _{i}$'s are Lovelock coefficients with dimensions
$(\mathrm{length})^{2i-2}$ and \cite{Lov}
\begin{equation}
\mathcal{L}_{i}=\frac{1}{2^{i}}\delta _{\nu _{1}\text{ }\nu _{2}\cdots \nu
_{2i}}^{\mu _{1}\mu _{2}\cdots \mu _{2i}}R_{\mu _{1}\mu _{2}}^{%
\phantom{\mu_1\mu_2}{\nu_1\nu_2}}\cdots R_{\mu _{2i-1}\mu _{2i}}^{%
\phantom{\mu_{2i-1} \mu_{2i}}{\nu_{2i-1} \nu_{2i}}}.  \label{LoveLag}
\end{equation}
A key property of this action is that the term proportional to $\alpha _{k}$
contributes to the equations of motion in dimensions with $n\geq 2k$. Hence
the above action with interaction terms quartic in the curvature tensor
contribute to the equations of motion only in nine and higher dimensions and
hence will not contribute in the desired five dimensions.

While Lovelock's Lagrangian yields second-order equations of motion for an
arbitrary spacetime, we limit ourselves to the case of spherically symmetric
spacetimes. The metric of $5$-dimensional spherically symmetric spacetime
may be written as
\begin{equation}
ds^{2}=-N^{2}(r)f(r)dt^{2}+\frac{dr^{2}}{f(r)}+r^{2}d\Sigma _{k,3}^{2},
\label{metric}
\end{equation}
where $d\Sigma _{k,3}^{2}$ represents the metric of a $3$-dimensional
hypersurface with constant curvature $6k$ and volume $V_{3}$. The first
three terms in the action (\ref{Act1}) contribute to the field equation in
five dimensions, while the third and fourth-order Lovelock terms do not.

Our aim is to include in the action terms quartic in the curvature that
contribute to the field equations in five dimensions and yield second-order
equations of motion for spherically symmetric spacetimes. We find that this
action may be written as
\begin{equation}
I_{\mathrm{G}}=\frac{1}{16\pi }\int d^{n+1}x\sqrt{-g}[-2\Lambda +\mathcal{L}%
_{1}+\mu _{2}\mathcal{L}_{2}+\mu _{3}\mathcal{X}_{3}+\mu _{4}\mathcal{X}_{4}%
],  \label{Act2}
\end{equation}
where $\mathcal{L}_{1}={R}$ is just the Einstein-Hilbert Lagrangian, $%
\mathcal{L}_{2}=R_{abcd}{R}^{abcd}-4{R}_{ab}{R}^{ab}+{R}^{2}$ is the second-order Lovelock (Gauss-Bonnet) Lagrangian, and $\mathcal{X}_{3}$\ is the
curvature-cubed Lagrangian
\begin{eqnarray}
\mathcal{X}_{3} &=&R_{abcd}R^{bedf}R_{e\,\,\,f}^{\,\,a\,\,\,\,\,c}+\frac{1}{%
(2n-1)(n-3)}\left( \frac{3(3n-5)}{8}R_{abcd}R^{abcd}R\right.  \notag \\
&&-3(n-1)R_{abcd}R^{abc}{}_{e}R^{de}+3(n+1)R_{abcd}R^{ac}R^{bd}  \notag \\
&&\left. +\,6(n-1)R_{ab}R^{bc}R_{c}{}^{a}-\frac{3(3n-1)}{2}R_{ab}R^{ab}R+%
\frac{3(n+1)}{8}R^{3}\right)  \label{X3}
\end{eqnarray}
obtained previously \cite{Myers1}.

In eight dimensions, there are 26 distinct scalar functions that are quartic
in the curvature tensor \cite{Fulling}. However, one may construct the fourth-order Lagrangian of
Lovelock gravity by combining the following 25 terms
\begin{eqnarray*}
&&R^{4},\text{ \ \ \ \ \ \ \ \ \ \ \ \ \ \ \ \ \ \ \ \ \ \ \ \ \ \ \ \ \ \ \
}R^{2}R_{ab}{}R{}^{ab},\text{ \ \ \ \ \ \ \ \ \ \ \ \ \ \ \ \ \ \ }%
R^{2}R_{abcd}{}R{}^{abcd}, \\
&&RR_{b}{}^{a}R_{c}{}^{b}R_{a}{}^{c},\text{ \ \ \ \ \ \ \ \ \ \ \ \ \ \ \ }%
RR_{c}{}^{a}R_{d}{}^{b}R_{ab}{}^{cd},\text{ \ \ \ \ \ \ \ \ \ \ \ }%
RR_{b}{}^{a}R_{de}{}^{bc}R_{ac}{}^{de}, \\
&&RR_{cd}{}^{ab}R_{ef}{}^{cd}R_{ab}{}^{ef},\text{ \ \ }\ \ \ \ \ \
RR_{ce}{}^{ab}R_{af}{}^{cd}R_{bd}{}^{ef},\text{ \ \ \ \ \ \ \ }%
R_{b}{}^{a}R_{a}{}^{b}R_{d}{}^{c}R_{c}{}^{d}, \\
&&R_{b}{}^{a}R_{c}{}^{b}R_{d}{}^{c}R_{a}{}^{d},\text{ \ \ \ \ \ \ \ \ \ \ \ }%
R_{b}{}^{a}R_{d}{}^{b}R_{e}{}^{c}R_{ac}{}^{de},\text{ \ \ \ \ \ \ \ \ }%
R_{b}{}^{a}R_{a}{}^{b}R_{ef}{}^{cd}R_{cd}{}^{ef}, \\
&&R_{b}{}^{a}R_{c}{}^{b}R_{ef}{}^{cd}R_{ad}{}^{ef},\text{ \ \ \ \ \ \ \ }%
R_{c}{}^{a}R_{d}{}^{b}R_{ef}{}^{cd}R_{ab}{}^{ef},\text{ \ \ \ \ \ \ \ }%
R_{c}{}^{a}R_{e}{}^{b}R_{af}{}^{cd}R_{bd}{}^{ef}, \\
&&R_{c}{}^{a}R_{e}{}^{b}R_{bf}{}^{cd}R_{ad}{}^{ef},\text{ \ \ \ \ \ \ \ }%
R_{b}{}^{a}R_{ad}{}^{bc}R_{fg}{}^{de}R_{ce}{}^{fg},\text{ \ \ \ \ }%
R_{b}{}^{a}R_{de}{}^{bc}R_{fg}{}^{de}R_{ac}{}^{fg}, \\
&&R_{b}{}^{a}R_{df}{}^{bc}R_{ag}{}^{de}R_{ce}{}^{fg},\text{ \ \ \ \ }%
(R^{abcd}R_{abcd})^2,\text{ \ \ \ \ \ \ \ \ \ \ \ \ }%
R{}^{abcd}R_{abc}{}^{e}R^{fgh}{}_{d}R_{fghe}, \\
&&R^{abcd}R_{efcd}R^{efgh}R_{abgh},\text{ \ }%
R_{cd}{}^{ab}R_{eg}{}^{cd}R_{ah}{}^{ef}R_{bf}{}^{gh},\text{ \ }%
R_{ce}{}^{ab}R_{ag}{}^{cd}R_{bh}{}^{ef}R_{df}{}^{gh}, \\
&&R_{ce}{}^{ab}R_{ag}{}^{cd}R_{dh}{}^{ef}R_{bf}{}^{gh}.
\end{eqnarray*}
in a particular way \cite{AYale}.

Since the $\chi _{3}$ Lagrangian contains no derivatives of the curvature
tensor, we shall construct the $\chi _{4}$ term in the action using only the
above 25 terms. For the metric (\ref{metric}) the function $N(r)$ performs
the role of the lapse function, making it possible to write the action as a
functional of $f(r)$ and its derivatives, with $N(r)$ appearing linearly in
the action.

Since the Riemann tensor has at most 2 derivatives of the metric functions
we find that there are at most 8 derivatives in any term for the
quartic-curvature action. We require all terms in the Lagrangian to vanish
that have more than two derivatives. For the metric ansatz (\ref{metric})
not all 25 terms above are needed to ensure the resultant equations of
motion are second-order differential equations. Remarkably we find that we
can choose
\begin{eqnarray}
\mathcal{X}_{4}\hspace{-0.2cm} &=&\hspace{-0.2cm}c_{1}R_{abcd}R^{cdef}R_{%
\phantom{hg}{ef}%
}^{hg}R_{hg}{}^{ab}+c_{2}R_{abcd}R^{abcd}R_{ef}R^{ef}+c_{3}RR_{ab}R^{ac}R_{c}{}^{b}+c_{4}(R_{abcd}R^{abcd})^{2}
\notag \\
&&\hspace{-0.1cm}%
+c_{5}R_{ab}R^{ac}R_{cd}R^{db}+c_{6}RR_{abcd}R^{ac}R^{db}+c_{7}R_{abcd}R^{ac}R^{be}R_{%
\phantom{d}{e}}^{d}+c_{8}R_{abcd}R^{acef}R_{\phantom{b}{e}}^{b}R_{%
\phantom{d}{f}}^{d}  \notag \\
&&\hspace{-0.1cm}%
+c_{9}R_{abcd}R^{ac}R_{ef}R^{bedf}+c_{10}R^{4}+c_{11}R^{2}R_{abcd}R^{abcd}+c_{12}R^{2}R_{ab}R^{ab}
\notag \\
&&\hspace{-0.1cm}%
+c_{13}R_{abcd}R^{abef}R_{ef}{}_{g}^{c}R^{dg}+c_{14}R_{abcd}R^{aecf}R_{gehf}R^{gbhd},
\label{X4}
\end{eqnarray}
without loss of generality. We must then choose the coefficients $c_{i}$ to
yield only a second-order contribution to the field equations. We find that
\begin{eqnarray}
c_{1} &=&-1404,\text{ \ \ \ \ }c_{2}=1848,\text{ \ \ \ \ \ }c_{3}=-25536,%
\text{ \ \ \ }c_{4}=-7422,\text{ \ \ \ \ }c_{5}=24672,  \notag \\
c_{6} &=&-5472,\text{ \ \ }c_{7}=77184,\text{ \ }c_{8}=-85824{,}\text{ \ \ \
}c_{9}=-41472{,}\text{ \ \ }c_{10}=-690,  \notag \\
c_{11} &=&1788,\text{ \ \ }c_{12}=6936,\text{ \ }c_{13}=7296{,}\text{ \ \ }%
c_{14}=42480  \label{C7}
\end{eqnarray}
is the unique solution up to a term proportional to the quartic Lovelock
Lagrangian.

Defining the dimensionless parameters $\hat{\mu}_{0}$ .... $\hat{\mu}_{4}$
to be
\begin{equation}
\hat{\mu}_{0}=-\frac{l^{2}}{6}\Lambda ,\text{ \ \ \ }\hat{\mu}_{2}=\frac{2}{%
l^{2}}\mu _{2},\hspace{0.5cm}\hat{\mu}_{3}=\frac{4}{7l^{4}}\mu _{3}\,\hspace{%
0.5cm}\hat{\mu}_{4}=\frac{21024}{l^{6}}\mu _{4}
\end{equation}
and integrating by parts, we find that the action (\ref{Act2}) per unit
volume $V_{3}$ reduces to the rather simple form
\begin{equation}
I_{\mathrm{G}}=\frac{3}{16\pi l^{2}}\int dtdr{N(r)}\left\{ r^{4}\left( \hat{%
\mu}_{0}+\psi +\hat{\mu}_{2}\psi ^{2}+\hat{\mu}_{3}\psi ^{3}+\hat{\mu}%
_{4}\psi ^{4}\right) \right\} ^{\prime },  \label{Igfin}
\end{equation}
where prime denotes the derivative with respect to $r$ and $\psi
=l^{2}r^{-2}(k-f)$.

\section{Generalization to $n+1$ Dimensions}

In this section we consider the action (\ref{Act2}) in $n+1$ dimensions for
the spherical metric
\begin{equation}
ds^{2}=-N^{2}(r)f(r)dt^{2}+\frac{dr^{2}}{f(r)}+r^{2}d\Sigma _{k,n-1}^{2},
\label{met2}
\end{equation}
where $d\Sigma _{k,n-1}^{2}$ represents the metric of an $(n-1)$-dimensional
hypersurface with constant curvature $(n-1)(n-2)k$ and volume $V_{n-1}$.
Using the same procedure as in the preceding section for five dimensions, we
can obtain the coefficients $c_{i}$'s in Eq. (\ref{X4}). The results are
somewhat cumbersome so we list them in the Appendix.

As before, we find that after integrating by parts and defining the
dimensionless parameters $\hat{\mu}_{0}$, $\hat{\mu}_{2}$, $\hat{\mu}_{3}$
and $\hat{\mu}_{4}$ to be
\begin{eqnarray*}
\hat{\mu}_{0} &=&-\frac{2l^{2}}{n(n-1)}\Lambda ,\text{ \ }\hat{\mu}_{2}=%
\frac{(n-2)(n-3)}{l^{2}}\mu _{2},\text{ \ \ \ }\hat{\mu}_{3}=\frac{%
(n-2)(n-5)(3n^{2}-9n+4)}{8(2n-1)l^{4}}\mu _{3}, \\
\hat{\mu}_{4} &=&{\frac{n\left( n-1\right) \left( n-2\right) ^{2}\left(
n-3\right) \left( n-7\right) ({{n}^{5}-15\,{n}^{4}+72\,{n}^{3}-156\,{n}%
^{2}+150\,n-42)}}{{l}^{6}}}\mu _{4},
\end{eqnarray*}
the action per unit volume reduces to
\begin{equation}
I_{\mathrm{G}}=\frac{(n-1)}{16\pi l^{2}}\int dtdrN(r)\left[ r^{n}(\hat{\mu}%
_{0}+\psi +\hat{\mu}_{2}\psi ^{2}+\hat{\mu}_{3}\psi ^{3}+\hat{\mu}_{4}\psi
^{4})\right] ^{\prime },  \label{Igfin2}
\end{equation}
where again $\psi =l^{2}r^{-2}(k-f)$. Note that in the absence of a
cosmological constant $\hat{\mu}_{0}=0$, while in the presence of a
positive/negative cosmological constant we take $\hat{\mu}_{0}=\pm 1$.

We pause to comment that $\hat{\mu}_{4}$ is zero in $8$ dimensions,
suggesting that $\mathcal{X}_{4}$ yields another topological invariant in
eight dimensions besides the 8-dimensional Euler density (given by $L_{4}$
in Eq. (\ref{LoveLag})). However it is straightforward to show that $%
\mathcal{X}_{4}$ has $8$-th order derivative terms for nontrivial $8$%
-dimensional geometries and therefore is not a topological invariant. Hence
we refer to this theory of gravity as quartic quasitopological gravity.
Note that our construction does not yield a nontrivial quartic interaction
term in $n+1\leq 4$ as well.

Varying the action (\ref{Igfin}) with respect to $N(r)$, we obtain
\begin{equation}
\left[ r^{n}(\hat{\mu}_{0}+\psi +\hat{\mu}_{2}\psi ^{2}+\hat{\mu}_{3}\psi
^{3}+\hat{\mu}_{4}\psi ^{4})\right] ^{\prime }=0  \label{Efr}
\end{equation}
for the equations of motion.  Formally this equation is the same as that obtained from 9-dimensional fourth-order Lovelock gravity in the spherically symmetric case. However the power of $r$ differs in (\ref{Efr}) from this case, since $n \geq 5$ can have any integer value except $8$.  The black hole solutions to this equation will consequently have analogous properties.  They will be asymptotically flat, AdS, or dS depending on the choice of parameters (as we will discuss below) and they will have a scalar curvature singularity at $r=0$ cloaked by an event horizon.

The solutions (\ref{Efr}) are the real roots of
the following quartic equation
\begin{equation}
\psi ^{4}+\frac{\hat{\mu}_{3}}{\hat{\mu}_{4}}\psi ^{3}+\frac{\hat{\mu}_{2}}{%
\hat{\mu}_{4}}\psi ^{2}+\frac{1}{\hat{\mu}_{4}}\psi +\frac{1}{\hat{\mu}_{4}}%
\kappa =0,  \label{Eq4}
\end{equation}
where
\begin{equation}
\kappa =\hat{\mu}_{0}-\frac{m}{r^{n}},  \label{kap}
\end{equation}
and $m$ is an integration constant which is related to the mass of the
spacetime.

The geometrical mass of black hole solutions is
\begin{equation}
m=\left( \hat{\mu}_{0}+k\frac{l^{2}}{r_{+}^{2}}+\hat{\mu}_{2}{k}^{2}\frac{%
l^{4}}{r_{+}^{4}}+\hat{\mu}_{3}{k}^{3}\frac{l^{6}}{r_{+}^{6}}+\hat{\mu}_{4}{k%
}^{4}\frac{l^{8}}{r_{+}^{8}}\right) r_{+}^{n}  \label{mh}
\end{equation}
in terms of the horizon radius $r_{+}$. Before considering the properties of
particular solutions, we compute the Hawking temperature
\begin{equation}
T=\frac{1}{4}\,{\frac{n\hat{\mu}_{0}r_{+}^{8}+\left( n-2\right) k{l}%
^{2}r_{+}^{6}+\left( n-4\right) {k}^{2}\hat{\mu}_{2}l^{4}r_{+}^{4}+\left(
n-6\right) k^{3}\hat{\mu}_{3}{l}^{6}r_{+}^{2}+\left( n-8\right) {k}^{4}\hat{%
\mu}_{4}{l}^{8}}{\left( \,r_{+}^{6}+2k\hat{\mu}_{2}{l}^{2}r_{+}^{4}\,+3k^{2}%
\hat{\mu}_{3}{l}^{4}r_{+}^{2}\,+4\hat{\mu}_{4}k^{3}{l}^{6}\right) \pi \,{l}%
^{2}r_{+}}}  \label{Temp}
\end{equation}
for the general black hole solution given by Eq. (\ref{chi4}). Clearly, $T$
always positive for $k=0$, and therefore there is no extreme black hole.
However, for $k=\pm 1$, extremal black hole solutions exist with horizon
radius $r_{\mathrm{ext}}$, where $r_{\mathrm{ext}}$ is the largest real root
of
\begin{equation}
n\hat{\mu}_{0}r_{\mathrm{ext}}^{8}+\left( n-2\right) k{l}^{2}r_{\mathrm{ext}%
}^{6}+\left( n-4\right) {k}^{2}\hat{\mu}_{2}l^{4}r_{\mathrm{ext}}^{4}+\left(
n-6\right) k^{3}\hat{\mu}_{3}{l}^{6}r_{\mathrm{ext}}^{2}+\left( n-8\right) {k%
}^{4}\hat{\mu}_{4}{l}^{8}.  \label{rext}
\end{equation}
Equation (\ref{rext}) has at least one real solution in the absence of a
cosmological constant ($\hat{\mu}_{0}=0$). Hence there exist black holes
with inner and outer horizons, extreme black holes or naked singularities,
depending on the choice of parameters.

However for nonzero cosmological constant, extreme black holes appear as
solutions provided
\begin{equation*}
\Delta =A^{3}+\frac{B^{2}}{2}>0,
\end{equation*}
where
\begin{eqnarray*}
A &=&3\left( n-2\right) \left( n-6\right) \hat{\mu}_{{3}}-12n\left(
n-8\right) \hat{\mu}_{{0}}\hat{\mu}_{{4}}-\,\left( n-4\right) ^{2}{\hat{\mu}%
_{{2}}^{2},} \\
B &=&{-9}\,\left( n-4\right) {\hat{\mu}_{{2}}}\left[ 8n\left( n-8\right)
\hat{\mu}_{{0}}\hat{\mu}_{{4}}+\left( n-2\right) \left( n-6\right) \hat{\mu}%
_{{3}}\right] \\
&&+27(n-2)^{2}\left( n-8\right) \hat{\mu}_{{4}}+27n\left( n-6\right) ^{2}%
\hat{\mu}_{{0}}\hat{\mu}_{{3}}^{2}+2\,\left( n-4\right) ^{3}{\hat{\mu}_{{2}%
}^{3}}.
\end{eqnarray*}
The mass of the extreme black hole may be obtained by using Eq. (\ref{mh})
and computing $m_{\mathrm{ext}}=m(r_{\mathrm{ext}})$. Then, our solution
corresponds to a black hole with inner and outer horizons provided $m>m_{\mathrm{%
ext}}$, an extreme black hole if $m=m_{\mathrm{ext}}$, and a naked
singularity for $m<m_{\mathrm{ext}}$

\section{Special Solutions}

We consider in this section special solutions of Eq. (\ref{Eq4}).
Eliminating the cubic term in Eq. (\ref{Eq4}) by use of the transformation
\begin{equation}
\psi =\chi -\frac{\hat{\mu}_{3}}{4\hat{\mu}_{4}}.
\end{equation}
yields
\begin{equation}
\chi ^{4}-\alpha \chi ^{2}+\beta \chi -\gamma =0,  \label{chi4}
\end{equation}
where
\begin{eqnarray}
\alpha &=&\frac{3\hat{\mu}_{3}^{2}}{8\hat{\mu}_{4}^{2}}-\frac{\hat{\mu}_{2}}{%
\hat{\mu}_{4}},\hspace{0.5cm}\beta =\frac{\hat{\mu}_{3}^{3}}{8\hat{\mu}%
_{4}^{3}}-\frac{\hat{\mu}_{2}\hat{\mu}_{3}}{2\hat{\mu}_{4}^{2}}+\frac{1}{%
\hat{\mu}_{4}},  \notag \\
\gamma &=&\frac{3\hat{\mu}_{3}^{4}}{256\hat{\mu}_{4}^{4}}-\frac{\hat{\mu}_{2}%
\hat{\mu}_{3}^{2}}{16\hat{\mu}_{4}^{3}}+\frac{\hat{\mu}_{3}}{4\hat{\mu}%
_{4}^{2}}-\frac{\hat{\mu}_{0}}{\hat{\mu}_{4}}+\frac{m}{\hat{\mu}_{4}r^{n}}.
\end{eqnarray}
The most general solution of (\ref{chi4}) will yield the most general
 metric solution for quartic quasitopological
gravity with constant curvature horizons. However special solutions will emerge for particular choices of the
coefficients; we first examine these.

\subsection{$\protect\alpha =0$, $\protect\beta =0$:}

In this case $\hat{\mu}_{3}$ and $\hat{\mu}_{4}$ are:
\begin{equation*}
{\hat{\mu}_{3}}=\frac{4{\hat{\mu}_{2}}^{2}}{9}\quad {\hat{\mu}_{4}}=\frac{2{%
\hat{\mu}_{2}}^{3}}{27}
\end{equation*}
and the asymptotically AdS solution is
\begin{equation}
f(r)=k+\frac{3}{2\hat{\mu}_{2}}\frac{r^{2}}{l^{2}}\left\{ 1\pm \left( 1-%
\frac{8{\hat{\mu}}_{2}}{3}\left[ \hat{\mu}_{0}-\frac{m}{r^{n}}\right]
\right) ^{1/4}\right\} .  \label{Fr1}
\end{equation}
Requiring nonsingular real solutions implies that $0<\hat{\mu}_{2}<3\hat{\mu%
}_{0}/8$.

\begin{figure}[h]
\centering {\includegraphics[width=12cm]{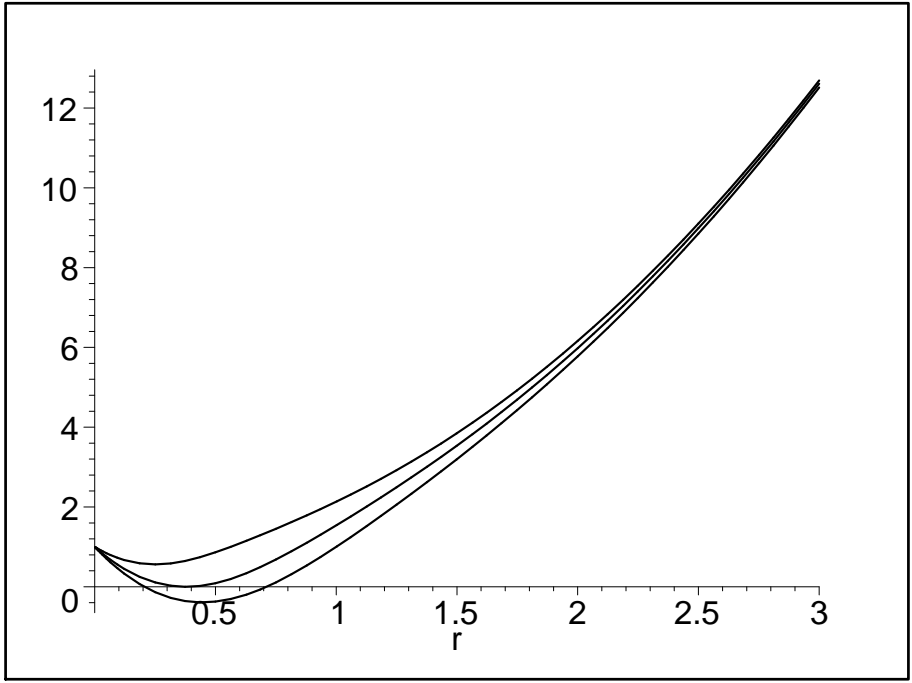}}
\caption{The asymptotically anti de Sitter case (A): $f(r)$ vs $r$ for $k=1$, $n=4$, $\hat{\protect\mu}_{0}=1$, $\hat{\protect\mu}_{2}=.2$, l=1 and $m<m_{\mathrm{ext}}$, $m=m_{\mathrm{ext}}$ and $m>m_{\mathrm{ext}}$ from up to
down, respectively.}
\label{FspAdS}
\end{figure}

For $\hat{\mu}_{0}=1$, the minus branch corresponds to an asymptotically AdS
black hole solution with two horizons provided $m>m_{\mathrm{ext}}$, an
extreme black hole if $m=m_{\mathrm{ext}}$, and a naked singularity for $%
m<m_{\mathrm{ext}}$ (see Fig. \ref{FspAdS}).
The plus branch always yields a naked singularity for $k=0,1$. However for $%
k=-1$ it corresponds to a black hole with a single horizon. The event
horizon is located at
\begin{equation}
\hat{\mu}_{0}x^{n+4}-x^{n+2}+{\hat{\mu}_{2}}x^{n}-\frac{4{\hat{\mu}_{2}}^{2}%
}{9}x^{n-2}+\frac{2{\hat{\mu}_{2}}^{3}}{27}x^{n-4}-ml^{-n}x^{4}=0
\label{ehpls}
\end{equation}
where $x=r_{+}/l$. In this case the mass parameter can even be negative
above a certain lower bound \cite{negblack}. A similar situation holds for $%
\hat{\mu}_{0}=-1$.

Although such solutions do not have a smooth general relativistic limit
as $\hat{\mu}_{2} \to 0$, it is possible that phase transitions to this branch from the minus branch
could occur.  This phenomenon has been demonstrated to take place in Gauss-Bonnet gravity
\cite{arXiv:0807.2864}.  Despite both branches having positive mass \cite{hep-th/0205318}, the plus
branch is perturbatively unstable. Quantum transitions can occur between the two vacua, and
neither the empty Einstein vacuum, nor the empty Gauss- Bonnet vacuum provide a good description of the stable quantum vacuum, since each becomes populated with bubbles of the other \cite{arXiv:0807.2864}.  Whether or not
a similar phenomenon takes place in quasitopological gravity remains an interesting topic for future investigation.
With this in mind, we will henceforth consider only the minus branch of the solutions.

\begin{figure}[h]
\centering {\includegraphics[width=12cm]{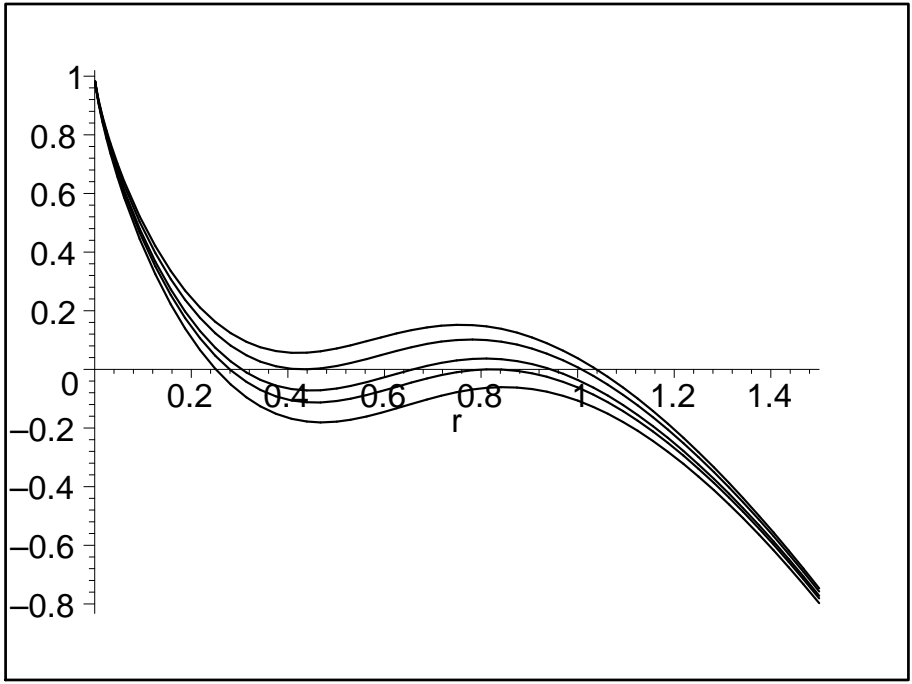}}
\caption{The asymptotically de Sitter case (A): $f(r)$ vs $r$ for $k=1$, $n=5$, $\hat{\protect\mu}_{0}=-1$, $\hat{\protect\mu}_{2}=.4$, l=1 and $m<m_{\mathrm{ext}}$, $m=m_{\mathrm{ext}}$, $m_{\mathrm{ext}}<m<m_{\mathrm{crit}}$
,$m=m_{\mathrm{crit}}$, and $m>m_{\mathrm{crit}}$ from up to down,
respectively.}
\label{FdS1}
\end{figure}

\begin{figure}[h]
\centering {\includegraphics[width=12cm]{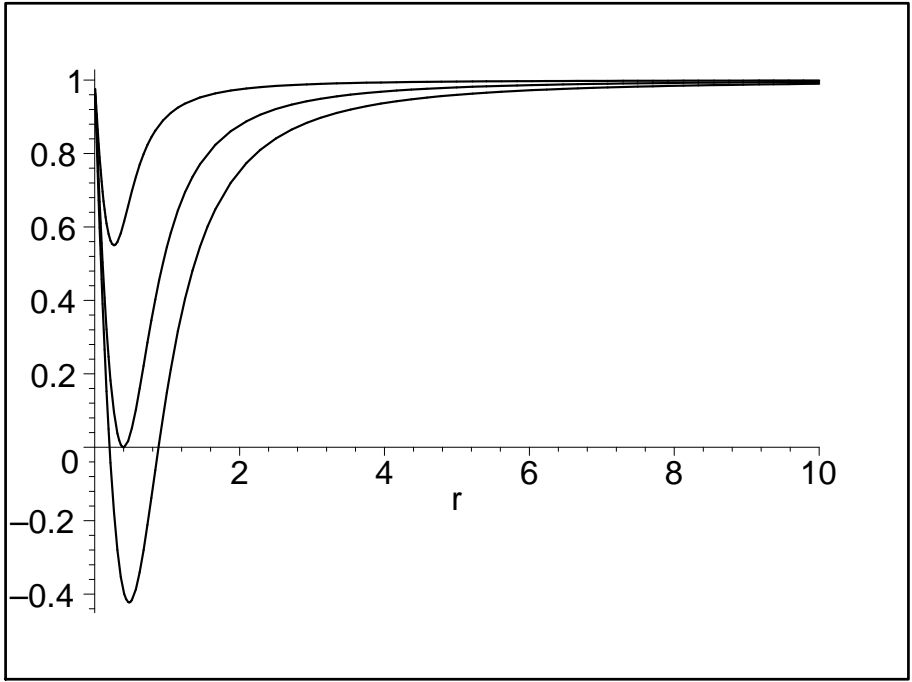}}
\caption{The asymptotically flat case (A) :$f(r)$ vs $r$ for $k=1$, $n=4$, $\hat{\protect\mu }_{0}=0$, $\hat{\protect\mu }_{2}=.5$, l=1 and $m<m_{\mathrm{ext}}$, $m=m_{\mathrm{ext}}$ and $m>m_{\mathrm{ext}}$ from up to
down, respectively. }
\label{FspFlat}
\end{figure}

For $\hat{\mu}_{0}=-1$ and $k=1$\ asymptotically de Sitter solutions are
present for the minus branch. These correspond to black holes with two
horizons, an extremal black hole with one horizon, or a naked singularity,
depending on the relative size of $m$. Asymptotically dS black holes exist
provided $m_{\mathrm{ext}}\leq m<m_{\mathrm{crit}}$, where $m_{\mathrm{ext}}$
and $m_{\mathrm{crit}}$ are the values of the mass parameter for the smaller
and larger root of $T=0$ respectively. We illustrate in Fig. \ref{FdS1}, the
behaviour of the metric function $f(r)$ for the various cases.
In the absence of a cosmological constant ($\hat{\mu }_{0}=0$), the case $%
k=1 $ yields an asymptotically flat black hole with metric function
\begin{equation}
f(r)=1+\frac{3}{2\hat{\mu }_{2}}\frac{r^{2}}{l^{2}}\left\{ 1-\left( 1+\frac{8%
{\hat{\mu }}_{2}}{3}\frac{m}{r^{n}}\right) ^{1/4}\right\}  \label{Fr2}
\end{equation}
This solution corresponds to a black hole with two horizons provided $m>m_{%
\mathrm{ext}}$, an extreme black hole if $m=m_{\mathrm{ext}}$, and a naked
singularity for $m<m_{\mathrm{ext}}$ (see Fig. \ref{FspFlat}).
\subsection{$\protect\beta =0$:}

\begin{figure}[h]
\centering {\includegraphics[width=12cm]{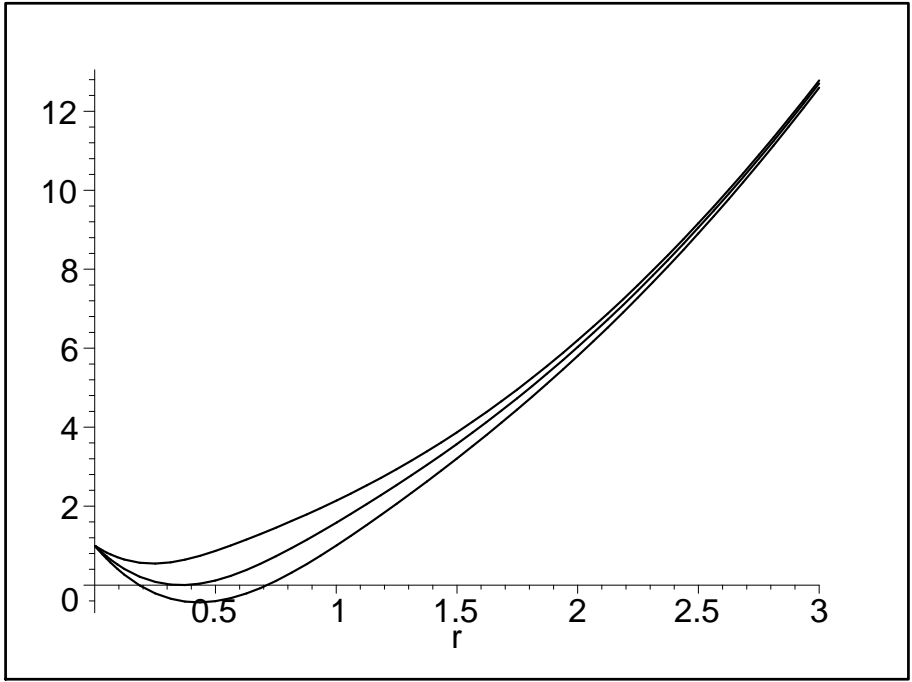}}
\caption{The asymptotically anti de Sitter case (B): $f(r)$ vs $r$ for $k=1$, $n=4$, $\hat{\protect\mu}_{0}=1$,$\hat{\protect\mu}_{2}=.2$,$\hat{\protect\mu}_{3}=.015$, l=1 and $m<m_{\mathrm{ext}}$, $m=m_{\mathrm{ext}}$ and $m>m_{\mathrm{ext}}$ from up to down, respectively.}
\label{Fsp2AdS}
\end{figure}

\begin{figure}[h]
\centering {\includegraphics[width=12cm]{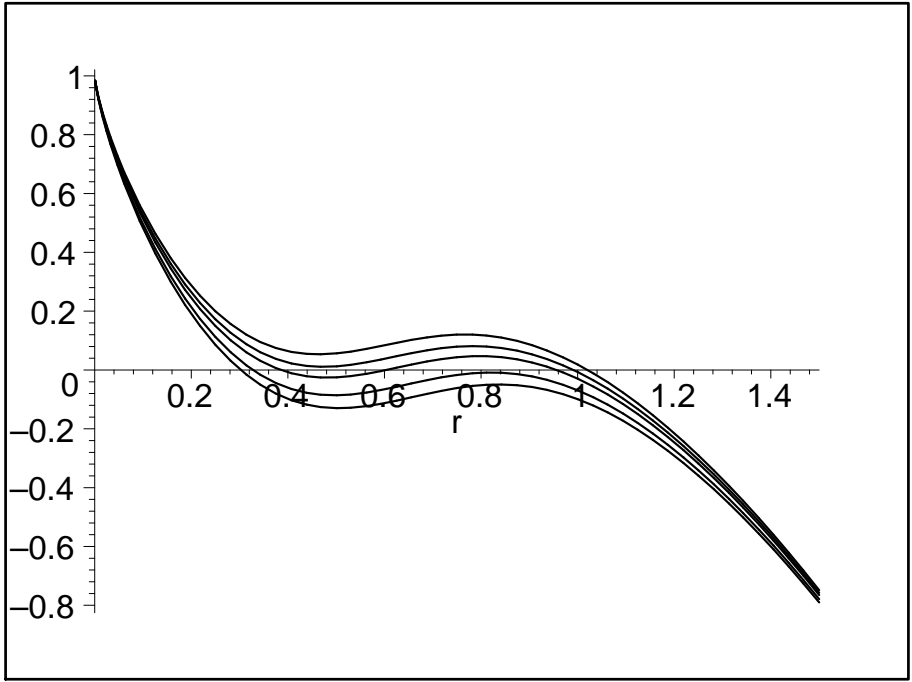}}
\caption{The asymptotically de Sitter case (B): $f(r)$ vs $r$ for $k=1$, $n=5$, $\hat{\protect\mu}_{0}=-1$, $\hat{\protect\mu}_{2}=.4$, $\hat{\protect\mu}_{3}=.08$, l=1 and $m<m_{\mathrm{ext}}$, $m=m_{\mathrm{ext}}$, $m_{\mathrm{ext}}<m<m_{\mathrm{crit}}$ ,$m=m_{\mathrm{crit}}$, and $m>m_{\mathrm{crit}}$ from up to down, respectively.}
\label{Fds2}
\end{figure}

Another special solution of Eq. (\ref{chi4}) corresponds to the case of $%
\beta =0$, for which Eq. (\ref{chi4}) is quadratic in $\chi ^{2}$ and
\begin{equation*}
\hat{\mu}_{2}={\frac{\hat{\mu}_{3}^{3}+8\hat{\mu}_{4}^{2}}{4\hat{\mu}_{3}%
\hat{\mu}_{4}}}.
\end{equation*}
The metric function $f(r)$ can be written as
\begin{equation}
f(r)=k+\frac{r^{2}}{l^{2}}\left( \frac{\hat{\mu}_{3}}{4\hat{\mu}_{4}}\pm
\sqrt{{\frac{\hat{\mu}_{3}^{3}-16\hat{\mu}_{4}^{2}}{16\hat{\mu}_{3}\hat{\mu}%
_{4}^{2}}+\sqrt{\frac{\hat{\mu}_{4}^{2}-\hat{\mu}_{0}\hat{\mu}_{3}^{2}}{\hat{%
\mu}_{3}^{2}\hat{\mu}_{4}^{2}}+\frac{m}{\hat{\mu}_{4}r^{n}}}}}\right) .
\label{Fr3}
\end{equation}
Since we are interested in black hole solutions, we choose the minus branch
of $f(r)$ for $k=0,1$. For $\hat{\mu}_{0}=1,$ the minus branch of this
solution corresponds to an asymptotically AdS black hole with two horizons,
an extreme black hole or a naked singularity provided $m>m_{\mathrm{ext}}$, $%
m=m_{\mathrm{ext}}$, and $m<m_{\mathrm{ext}}$, respectively. In Fig. \ref
{Fsp2AdS} we illustrate the various cases.
For $\hat{\mu}_{0}=-1$ and $k=1$, the solution corresponds
to an asymptotically de Sitter black hole with two horizons if $m_{\mathrm{%
ext}}<m<m_{\mathrm{crit}}$, an extremal black hole with one horizon if $m=m_{%
\mathrm{ext}}$, or a naked singularity otherwise. We illustrate the
different possibilities in Fig. \ref{Fds2}.

\begin{figure}[h]
\centering {\includegraphics[width=12cm]{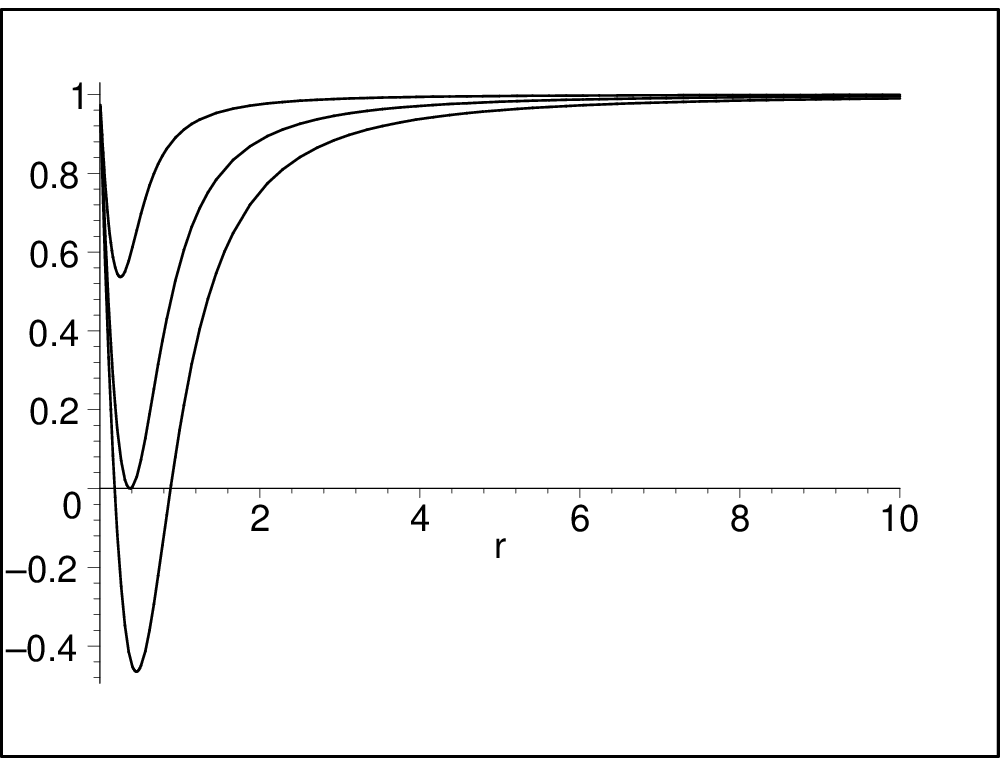}}
\caption{The asymptotically flat case (B): $f(r)$ vs $r$ for $k=1$, $n=4$, $\hat{\protect\mu}_{0}=0$,$\hat{\protect\mu}_{2}=.2$, $\hat{\protect\mu}_{3}=.015$, l=1 and $m<m_{\mathrm{ext}}$, $m=m_{\mathrm{ext}}$ and $m>m_{\mathrm{ext}}$ from up to down, respectively.}
\label{Fsp2Flat}
\end{figure}

For zero cosmological constant and $k=1$, the metric function for the
asymptotically flat black hole solution is shown in Fig. \ref{Fsp2Flat} for
various mass parameters.
 For $k=-1$ we also find black hole solutions with a
single horizon for the plus branch; we have not illustrated the metric
function here. We note, as per our earlier discussion, that this plus branch has no smooth Einsteinian limit.

\section{General Solutions}

We consider first asymptotically (A)dS solutions, for which we have two real
solutions provided $\Delta $ at infinity ($\kappa \to \hat{\mu}_{0}$) is
positive where
\begin{equation*}
\Delta =\frac{C^{3}}{27}+\frac{D^{2}}{4}
\end{equation*}
and
\begin{equation}
C={\frac{3\hat{\mu}_{3}-{\hat{\mu}_{2}}^{2}}{3{\hat{\mu}_{4}}^{2}}}-\,{\frac{%
4\kappa}{\hat{\mu}_{4}}}
\end{equation}
\begin{equation}
D={\frac{2}{27}}\,{\frac{{\hat{\mu}_{2}}^{3}}{{\hat{\mu}_{4}}^{3}}}-\frac{1}{%
3}\,\left( {\frac{\hat{\mu}_{3}}{{\hat{\mu}_{4}}^{2}}} + 8\,{\frac{\kappa}{%
\hat{\mu}_{4}}}\right) \frac{\hat{\mu}_{2}}{\hat{\mu}_{4}}+{\frac{{\hat{\mu}%
_{3}}^{2}\kappa}{{\hat{\mu}_{4}}^{3}}}+\frac{1}{{\hat{\mu}_{4}}^{2}}
\end{equation}
The real solutions of Eq. (\ref{Eq4}) are
\begin{equation}
f(r)=k+\frac{r^{2}}{l^{2}}\left( \frac{\hat{\mu}_{3}}{4\hat{\mu}_{4}}+\frac{1%
}{2}R\pm \frac{1}{2}E\right) .  \label{F4}
\end{equation}
where
\begin{eqnarray}
R &=&\left( \frac{{\hat{\mu}_{3}}^{2}}{4{\hat{\mu}_{4}}^{2}}-\frac{2\hat{\mu}%
_{2}}{3\hat{\mu}_{4}}+\left( {\frac{q}{2}+\sqrt{\Delta }}\right)
^{1/3}+\left( {\frac{q}{2}-\sqrt{\Delta }}\right) ^{1/3}\right) ^{1/2},
\label{RR} \\
E &=&\left( \frac{3{\hat{\mu}_{3}}^{2}}{4{\hat{\mu}_{4}}^{2}}-\frac{2\hat{\mu%
}_{2}}{\hat{\mu}_{4}}-R^{2}-\frac{1}{4R}\left[ \frac{4\hat{\mu}_{2}\hat{\mu}%
_{3}}{{\hat{\mu}_{4}}^{2}}-\frac{8}{\hat{\mu}_{4}}-\frac{{\hat{\mu}_{3}}^{3}%
}{{\hat{\mu}_{4}}^{3}}\right] \right) ^{1/2}  \label{EE}
\end{eqnarray}
describing the two physical branches of the solution.

\begin{figure}[h]
\centering {\includegraphics[width=12cm]{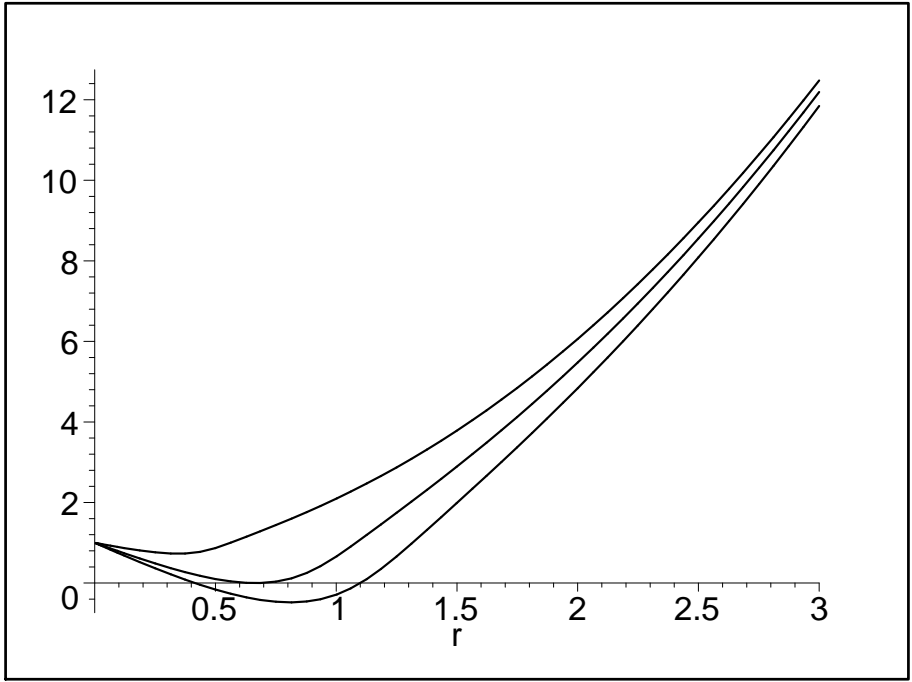}}
\caption{The general asymptotically anti de Sitter case: $f(r)$ vs $r$ for $k=1$, $n=4$, $\hat{\protect\mu}_{0}=1$, $\hat{\protect\mu}_{2}=.2$, $\hat{\protect\mu}_{3}=.1$,$\hat{\protect\mu}_{4}=.06$, l=1 and $m<m_{\mathrm{ext}}
$, $m=m_{\mathrm{ext}}$ and $m>m_{\mathrm{ext}}$from up to down,
respectively .}
\label{FAdS}
\end{figure}

Again we are interested in black hole solutions  that have a smooth Einsteinian limit. Therefore we choose the
minus branch of $f(r)$
\begin{equation}
f(r)=k+\frac{r^{2}}{l^{2}}\left( \frac{\hat{\mu}_{3}}{4\hat{\mu}_{4}}+\frac{1%
}{2}R-\frac{1}{2}E\right) .  \label{Fr4}
\end{equation}
Figure \ref{FAdS} shows the metric function $f(r)$ for different values of
mass parameters with $\hat{\mu}_{0}=+1$.
For $k=-1$ the solution yields a black hole with one horizon.

\begin{figure}[h]
\centering {\includegraphics[width=12cm]{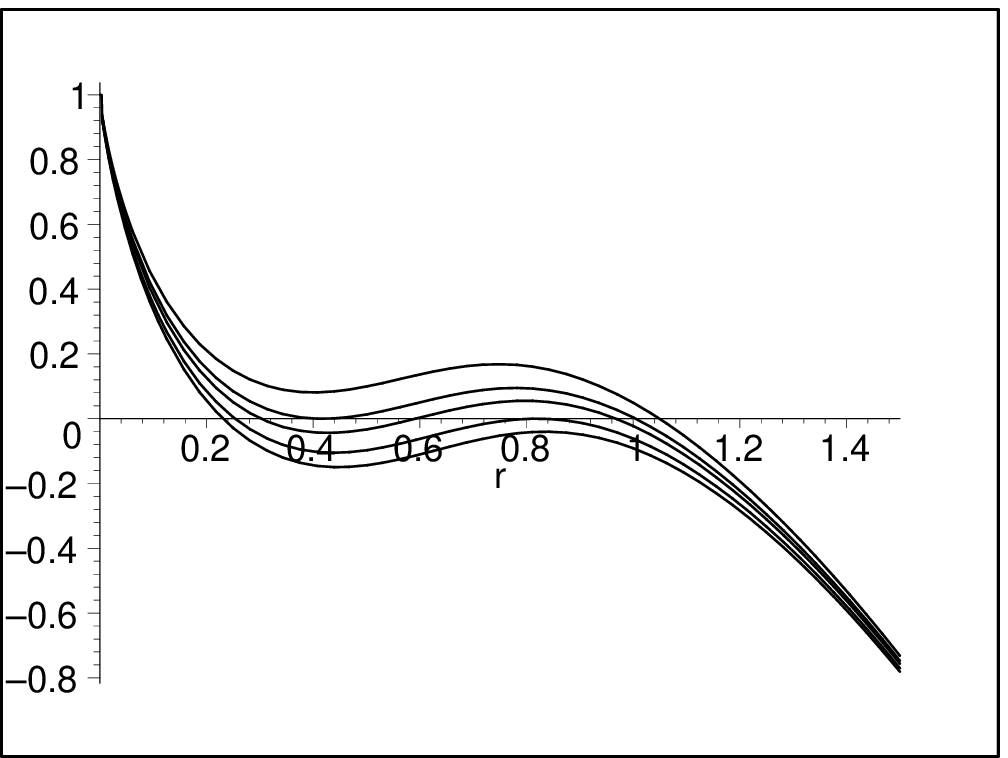}}
\caption{The general asymptotically de Sitter case: $f(r)$ vs $r$ for $k=1$, $n=5$, $\hat{\protect\mu}_{0}=-1$, $\hat{\protect\mu}_{2}=.4$, $\hat{\protect\mu}_{3}=.1$,$\hat{\protect\mu}_{4}=.002$, l=1 and $m<m_{\mathrm{ext}}$, $m=m_{\mathrm{ext}}$, $m_{\mathrm{ext}}<m<m_{\mathrm{crit}}$ $m=m_{\mathrm{crit}}$, and $m>m_{\mathrm{crit}}$ from up to down, respectively..}
\label{FdS3}
\end{figure}

For $\hat{\mu}_{0}=-1$ and $k=1$, the solution is that of an asymptotically
de Sitter black hole with two horizons if $m_{\mathrm{ext}}<m<m_{\mathrm{crit%
}}$, an extremal black hole with one horizon if $m=m_{\mathrm{ext}}$, or a
naked singularity otherwise (see Fig. \ref{FdS3}).

\begin{figure}[h]
\centering {\includegraphics[width=12cm]{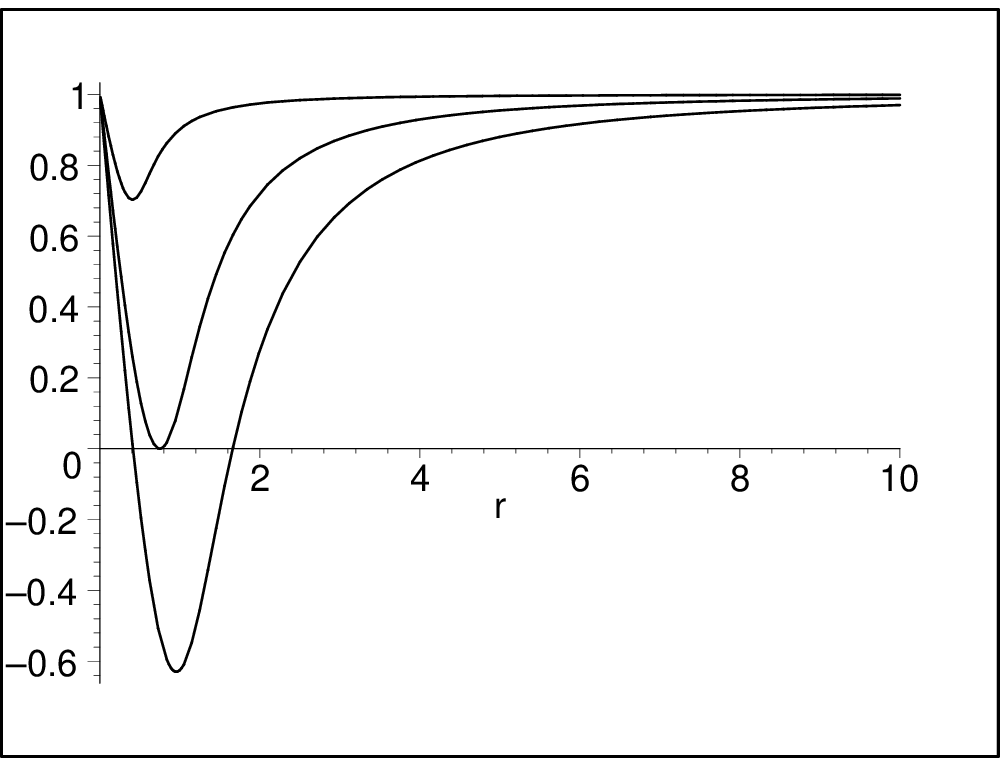}}
\caption{The general asymptotically flat case: $f(r)$ vs $r$ for $k=1$, 
$n=4$, $\hat{\protect\mu }_{0}=0$,$\hat{\protect\mu }_{2}=.2$, $\hat{\protect\mu }_{3}=.1$,$\hat{\protect\mu }_{4}=.06$, l=1 and $m<m_{\mathrm{ext}}$, $m=m_{\mathrm{ext}}$ and $m>m_{\mathrm{ext}}$ from up to down, respectively.}
\label{Fflat}
\end{figure}

Second, we consider asymptotically flat solutions. These are present only if
$\hat{\mu}_{0}=0$ (the cosmological constant vanishes\textbf{)}, implying
that Eq. (\ref{Eq4}) reduces to
\begin{equation}
\psi _{\infty }\left( \psi _{\infty }^{3}+\frac{\hat{\mu}_{3}}{\hat{\mu}_{4}}%
\psi _{\infty }^{2}+\frac{\hat{\mu}_{2}}{\hat{\mu}_{4}}\psi _{\infty }+\frac{%
1}{\hat{\mu}_{4}}\right) =0  \label{Eqflat}
\end{equation}
in the large-$r$ limit. We see from Eq. (\ref{Eqflat}) that one can have an
asymptotically flat solution with $\psi _{\infty }=0$. This asymptotically
flat solution can be written down by substituting $\hat{\mu}_{0}=0$ in Eq. (%
\ref{Fr4}). It corresponds to a black hole with inner and outer horizons
provided $m>m_{\mathrm{ext}}$, an extreme black hole if $m=m_{\mathrm{ext}}$%
, and a naked singularity for $m<m_{\mathrm{ext}} $ where $m_{\mathrm{ext}}$
can be calculated numerically. The metric functions for these black holes
are shown in Fig. \ref{Fflat}.

\section{Entropy Density}

The entropy of the black hole solutions can be calculated through the use of
the formula \cite{WaldEnt}
\begin{equation*}
S=-2\pi \oint d^{n-1}x\sqrt{\tilde{g}}\ \frac{\partial {\mathcal{L}}}{%
\partial R_{abcd}}\hat{\varepsilon}_{ab}\hat{\varepsilon}_{cd},
\end{equation*}
where $\mathcal{L}$ is the Lagrangian, $\tilde{g}$ is the determinant of the
induced metric on the horizon and $\hat{\varepsilon}_{ab}$ is the binormal
to the horizon. For the static black holes considered here
\begin{equation*}
Y=\frac{\partial {\mathcal{L}}}{\partial R_{abcd}}\hat{\varepsilon}_{ab}\hat{%
\varepsilon}_{cd}
\end{equation*}
is constant on the horizon and so the entropy density is $s=S/V_{n-1}=-2\pi r_{+}^{n-1}Y$. For Einstein Lagrangian, $Y_{1}=-1/(8\pi )$
and the resulting entropy density is the expected Bekenstein-Hawking entropy
$s=1/4r_{+}^{n-1}$ \cite{Haw}. Applying this formalism to
Gauss-Bonnet and cubic terms, one obtains \cite{Myers1}
\begin{eqnarray}
s_{2} &=&\frac{\hat{\mu}_{2}\,l^{2}r_{+}^{n-1}}{2(n-2)(n-3)}\left(
R-2\left( R_{\,\,t}^{t}+R_{\,\,r}^{r}\right) +2R_{\,\,\,\,tr}^{tr}\right)
\notag \\
&=&\frac{(n-1)l^{2}}{2(n-3)r_{+}^{2}}k\hat{\mu}_{2}r_{+}^{n-1} \\
s_{3} &=&\frac{4\,\hat{\mu}_{3}\,l^{4}r_{+}^{n-1}}{%
(n-2)(n-3)(n-5)(3n^{2}-9n+4)}\left[ \frac{3(n-3)(2n-1)}{2}\left(
R^{tm}{}_{tn}R^{rn}{}_{rm}-R^{tm}{}_{rn}R^{r}{}_{m}{}_{t}{}^{n}\right)
\right.  \notag \\
&&-\frac{3(n-1)}{n-3}\left(
R^{tr}{}_{tm}R_{r}{}^{m}-R^{tr}{}_{rm}R_{t}{}^{m}+\frac{1}{4}\left(
R_{mnpr}R^{mnpr}+R_{mnpt}R^{mnpt}\right) \right)  \notag \\
&&+\frac{3(3n-5)}{8(n-3)}\left( 2R\,R^{tr}{}_{tr}+\frac{1}{2}%
R_{mnpq}R^{mnpq}\right) +\frac{9(n-1)}{2(n-3)}\left(
R^{rm}R_{rm}+R^{tm}R_{tm}\right)  \notag \\
&&+\frac{3(n+1)}{2(n-3)}\left(
R^{t}{}_{t}R^{r}{}_{r}-R^{t}{}_{r}R^{r}{}_{t}+R^{r}{}_{mrn}R^{mn}+R^{t}{}_{mtn}R^{mn}\right)
\notag \\
&&\left. -\frac{3(3n-1)}{4(n-3)}\left( R_{mn}R^{mn}+R\left(
R^{r}{}_{r}+R^{t}{}_{t}\right) \right) +\frac{9(n+1)}{16(n-3)}R^{2}\right]
\notag \\
&=&\frac{3(n-1)l^{4}}{4(n-5)r_{+}^{4}}k^{2}\hat{\mu}_{3}r_{+}^{n-1}\,,
\end{eqnarray}
respectively. We can use the same formalism to obtain the entropy of the
quartic term (\ref{X4}). It is a matter of calculation to show that $Y_{4}$
reduces to
\begin{eqnarray}
Y_{4} &=&-\frac{1}{16\pi }\frac{\hat{\mu}_{4}l^{6}}{(n-7)n(n-1){(n-2)}%
^{2}(n-3)({{n}^{5}-15\,{n}^{4}+72\,{n}^{3}-156\,{n}^{2}+150\,n-42)}}[  \notag
\\
&&16c_{1}R^{rtef}R^{hg}{}_{ef}R_{hgrt}+2c_{2}(4R_{ef}R^{ef}R^{tr}{}_{tr}+R_{abcd}R^{abcd}(R^{r}{}_{r}+R^{t}{}_{t}))+c_{3}(2R_{ab}R^{ac}R_{c}{}^{b}
\notag \\
&&+3R(R^{ar}R_{ar}+R^{at}R_{at}))+16c_{4}R_{abcd}R^{abcd}R^{tr}{}_{tr}+4c_{5}(R_{ab}(R^{a}{}_{r}R^{br}+R^{a}{}_{t}R^{bt}))
\notag \\
&&+2c_{6}(R(R^{t}{}_{t}R^{r}{}_{r}-R^{t}{}_{r}R^{r}{}_{t})+RR^{ac}(R_{a}{}^{r}{}_{cr}+R_{a}{}^{t}{}_{ct})+R_{abcd}R^{ac}R^{bd})
\notag \\
&&+c_{7}(R^{r}{}_{r}R^{te}R_{te}+R^{t}{}_{t}R^{re}R_{re}-R^{t}{}_{r}R^{re}R_{te}-R^{r}{}_{t}R^{te}R_{re}+R^{be}R^{d}{}_{e}(R^{r}{}_{brd}+R^{t}{}_{btd})
\notag \\
&&+2R_{arcd}R^{ac}R^{dr}+2R_{atcd}R^{ac}R^{dt})+c_{8}(R_{r}{}^{ref}R_{te}R^{t}{}_{f}+R_{t}{}^{tef}R_{re}R^{r}{}_{f}-2R^{tref}R_{re}R_{tf}
\notag \\
&&+2R_{rbtd}(R^{br}R^{dt}-R^{bt}R^{dr})+2R^{d}{}_{e}(R_{arcd}R^{arce}+R_{atcd}R^{atce}))
\notag \\
&&+2c_{9}R_{ef}(R^{r}{}_{r}R_{t}{}^{etf}+R^{t}{}_{t}R_{r}{}^{erf}-2R_{tr}R^{retf}+R^{bedf}(R^{r}{}_{brd}+R^{t}{}_{btd}))
\notag \\
&&+8c_{10}R^{3}+8c_{11}(2R^{2}R^{tr}{}_{tr}+RR^{abcd}R_{abcd})  \notag \\
&&+2c_{12}(R^{2}(R^{r}{}_{r}+R^{t}{}_{t})+2RR_{ab}R^{ab})+c_{13}(4R^{abtr}(R_{abtd}R^{d}{}_{r}-R_{abrd}R^{d}{}_{t})+4R^{trcd}R_{trcg}R_{d}{}^{g}
\notag \\
&&+(R_{abcr}R_{ef}{}^{cr}+R_{abct}R_{ef}{}^{ct})R^{abef})+8c_{14}(R_{r}{}^{erf}R_{gehf}R^{gth}{}_{t}-R^{terf}R_{gehf}R^{g}{}_{r}{}^{h}{}_{t})],
\label{Y4}
\end{eqnarray}
where the $c_{i}$'s are given in Appendix. Now integrating over the horizon
and dividing by $V_{n-1}$, the entropy density reduces to
\begin{equation*}
s_{4}=\frac{(n-1)l^{6}}{(n-7)r_{+}^{6}}k^{3}\hat{\mu}%
_{4}r_{+}^{n-1}\,.
\end{equation*}
Combining all of these expressions, the entropy density for quartic
quasitopological gravity becomes
\begin{equation}
s=\frac{{r_{+}^{n-1}}}{4}\left( 1+2\,k\hat{\mu}_{2}{\frac{\left(
n-1\right) {l}^{2}}{\left( n-3\right) r_{+}^{2}}}+3k^{2}\hat{\mu}_{3}{\frac{%
\left( n-1\right) {\ l}^{4}}{\left( n-5\right) r_{+}^{4}}}+4{k}^{3}\hat{\mu}%
_{4}{\frac{\left( n-1\right) {l}^{6}}{\left( n-7\right) r_{+}^{6}}}\right)
\end{equation}
A simple method of finding the energy per unit volume $V_{n-1}$ is through
the use of first law of thermodynamics, $d\mathcal{M}=Tds$, which
gives \cite{WaldEnt}
\begin{equation}\label{massrplus}
\mathcal{M}=\int^{r_{+}}T\left( \frac{\partial s}{\partial r_{+}}%
\right) dr_{+}={\frac{\left( n-1\right) r_{+}^{n}}{16\pi \,}}\left( \hat{\mu}%
_{0}+k\frac{l^{2}}{r_{+}^{2}}+\hat{\mu}_{2}{k}^{2}\frac{l^{4}}{r_{+}^{4}}+%
\hat{\mu}_{3}{k}^{3}\frac{l^{6}}{r_{+}^{6}}+\hat{\mu}_{4}{k}^{4}\frac{l^{8}}{%
r_{+}^{8}}\right) .
\end{equation}
The energy density can be written in term of the geometrical mass by use of
Eq. (\ref{mh}) as
\begin{equation*}
\mathcal{M}={\frac{n-1}{16\pi \,}m.}
\end{equation*}

We pause to comment that for $k=-1$ it is possible to have negative mass
and/or entropy for certain values of the couplings. The phenomenon was
originally noted for the Einstein anti de Sitter case a number of years ago
\cite{negblack}. In this case the mass can be negative up to a certain
extremal value; the entropy is always positive. This situation also can
happen in Einstein-Gauss-Bonnet gravity. However the quasitopological terms
allow for both quantities to be negative for certain values of the
parameters (a situation that can also occur for third-order Lovelock gravity
\cite{Dehghani:2009zzb}). If the entropy is negative it is not clear what
solution is suitable as a reference state vacuum solution.

One approach to treating this problem is to add an overall constant to the entropy such
that $s \geq 0$ \cite{gr-qc/0402044}. This approach assumes that ${\cal M} \geq 0$, from which
a minimal value for $r_+$ is then obtained using (\ref{massrplus}). If $\hat{\mu}_{3}$ and $\hat{\mu}_{4}$
both vanish, such a minimal value is assured, and a minimal value of the entropy is obtained.  While having
the peculiar feature that the entropy vanishes despite the nonvanishing surface area of the minimal $r_+ = r_{+ min}$ black hole, an ambiguity in the Noether charge approach makes such an assignment possible \cite{gr-qc/0402044}.
However if  $\hat{\mu}_{3}\neq 0$ and/or $ \hat{\mu}_{4}\neq 0$  there will be several possible minima for $r_+$,
and it is no longer completely clear what assignment should be made in order to ensure the entropy remain positive.
This situation will also hold in all Lovelock theories third-order and higher, and we leave a complete treatment of
this subject for future study.

A general
thermodynamic treatment of these black holes can be carried out along lines
similar to that recently carried out for Lovelock black holes \cite{Camanho:2011rj}.

\section{Stability of the solutions}

An investigation of the full stability of the solutions we have obtained is beyond the scope of this paper. As a first step, we here  consider the stability of the solutions against a class of small
nonspherical  perturbations.

The metric of a slowly rotating solution in
five dimensions may be written as
\begin{equation}
ds^{2}=-f(r)dt^{2}+\frac{dr^{2}}{f(r)}+2ag(r)h(\theta )dtd\varphi
+r^{2}d\Omega ^{2},  \label{Slow}
\end{equation}
where $d\Omega ^{2}$ is the metric of a 3-sphere and $a$ is the rotation
parameter, which is assumed to be small. The first three terms of the action
are stable against a nonspherical perturbation, while the fourth one (the cubic quasitoplogical term)
is stable against   small nonspherical
perturbations \cite{Myers1}.

Here, we consider the stability of the solutions
 for the quartic quasitopological term against the above class of small nonspherical
perturbations. Consider the first scalar term in the Lagrangian (\ref
{X4}). We find for the slowly rotating
metric (\ref{Slow}) that
\begin{equation*}
R_{abcd}R^{abef}R_{efgh}R^{cdgh}=f^{\prime \prime 4}+\frac{6}{r^{4}}%
f^{\prime 4}+\frac{48}{r^{8}}[1-f(r)]^{2}+O(a^{2}),
\end{equation*}
where prime denotes the derivative with respect to the coordinate $r$. This
shows that a Lagrangian containing only the first term has no stable
solution against a nonspherical small perturbation. Upon further investigation we find this happens
for all the other terms of the quartic quasitopological term given in (\ref
{X4}).

Consider next the Lagrangian (\ref{X4}) with the coefficients given
in Eq. (\ref{C7}). To first order in $a$ we find that it reduces to
\begin{equation*}
\mathcal{X}_{4}=\frac{84096}{r^{8}}\left\{ (1-f)^{3}r^{2}f^{\prime \prime }-3%
\left[ (1-f)+rf^{\prime }\right] ^{2}\right\} +O(a^{2}).
\end{equation*}
and so the perturbed quartic Lagrangian contains  most second order derivatives. Therefore the field equation is a second order differential equation to linear order in $a$, and so
the spherical solutions are stable.

\section{Holographic hydrodynamics}

As a first step in understanding the role of our theory in the context of the AdS/CFT correspondence, we compute the ratio of shear
viscosity to entropy, $\eta /s$, leaving other subjects   such
as the holographic trace anomaly and holographic computation of energy fluxes for future study.

The first computations of $\eta /s$
from an AdS/CFT perspective appeared in \cite{Polic} for Einstein gravity, and
leading corrections  for
strongly coupled $N=4$ super-Yang-Mills theory subsequently followed \cite{Myers3,Buch2}.
These computations have been carried out for second \cite{Lg2} and
third-order Lovelock theories \cite{Lg3} and quasitopological gravity \cite{Myers2}.  Further
investigations also provided increasingly efficient techniques for these
calculations \cite{Iqbal, Pau2}. Here, we use the pole method  \cite{Pau2}, for the planar class of metrics
\begin{equation}
ds^{2}=\frac{r^{2}}{l^{2}}\left( \psi
(r)dt^{2}+dx_{1}^{2}+dx_{2}^{2}+dx_{3}^{2}\right) -\frac{l^{2}dr^{2}}{%
r^{2}\psi (r)},  \label{met3}
\end{equation}
where $\psi (r)$ is the root of Eq. (\ref{Eq4}) with $\hat{\mu }_{0}=1$.

Employing the transformation $z=1-r^{-2}m^{1/2}$, the metric (\ref{met3})
becomes
\begin{equation}
ds^{2}=\frac{m^{1/2}}{l^{2}(1-z)}\left( \psi
(z)dt^{2}+dx_{1}^{2}+dx_{2}^{2}+dx_{3}^{2}\right) -\frac{l^{2}dz^{2}}{%
4(1-z)\psi (z)},  \label{met4}
\end{equation}
where $\psi (z)$ has a simple zero at the horizon located at $z=0$. Thus $%
\psi (z)$ may be expanded as
\begin{equation}
\psi (z)=\psi _{0}^{(1)}z+\psi _{0}^{(2)}z^{2}+\psi _{0}^{(3)}z^{3}+\psi
_{0}^{(4)}z^{4}+...,  \label{Tay}
\end{equation}
where $\psi _{0}^{(i)}$ is the $i$th derivative of $\psi (z)$ at $z=0$.
Using Eq. (\ref{Eq4}) and the Taylor expansion (\ref{Tay}), the \ expansion
coefficients can be obtained as
\begin{eqnarray}
\psi _{0}^{(1)} &=&-2,\text{ \ \ \ \ \ }\psi _{0}^{(2)}=2(1-4\hat{\mu}_{2}),%
\text{ \ \ \ \ \ }\psi _{0}^{(3)}=24(\hat{\mu}_{2}-4\hat{\mu}_{2}^{2}+2\hat{%
\mu}_{3}),  \notag \\
\psi _{0}^{(4)} &=&24\left[ \hat{\mu}_{2}-24\hat{\mu}_{2}^{2}+80\hat{\mu}%
_{2}(\hat{\mu}_{3}-\hat{\mu}_{2}^{2})-12\hat{\mu}_{3}-16\hat{\mu}_{4}\right]
.  \label{Coef}
\end{eqnarray}

We perturb the metric (\ref{met4}) by the shift
\begin{equation}
dx_{i}\rightarrow dx_{i}+\varepsilon e^{-iwt}dx_{j},  \label{Pert}
\end{equation}
and we calculate the Lagrangian density. Because of the off-shell
perturbation (\ref{Pert}), there exists a pole at $z=0$ in the (otherwise)
on-shell action. The shear viscosity is \cite{Pau2}
\begin{equation*}
\eta =-8\pi T\lim_{\omega ,\varepsilon \rightarrow 0}\frac{Res_{z=0}%
\mathcal{L}}{\omega ^{2}\varepsilon ^{2}},
\end{equation*}
where $Res_{z=0}\mathcal{L}$ denotes the residue of the pole in the
Lagrangian density, and $T$ is the Hawking temperature give in Eq. (\ref
{Temp}) as $T=r_{+}(\pi l^{2})^{-1}=m^{1/4}(\pi l^{2})^{-1}$. It is a matter
of calculation to show that the shear viscosity reduces to
\begin{eqnarray*}
\eta  &=&\frac{m^{3/4}}{16\pi l^{3}}\{1-4\hat{\mu }_{2}-36\hat{\mu }%
_{3}\left( 9-64\hat{\mu}_{2}+128\hat{\mu}_{2}^{2}-48\hat{\mu}_{3}\right)  \\
&&-\frac{96}{73}\hat{\mu}_{4}\left( 1491-10800\hat{\mu}_{2}+28864\hat{\mu}%
_{2}^{2}-6240\hat{\mu}_{3}+10752\hat{\mu}_{2}\hat{\mu}_{3}-25088\hat{\mu}%
_{2}^{3}\right)\} .
\end{eqnarray*}
Now, using the fact that the entropy density of the black brane is $s=m^{3/4}(4l^{3})^{-1}$, the ratio of shear viscosity to entropy is
\begin{eqnarray*}
\frac{\eta }{s} &=&\frac{1}{4\pi }\{1-4\hat{\mu }_{2}-36%
\hat{\mu }_{3}\left( 9-64\hat{\mu}_{2}+128\hat{\mu}_{2}^{2}-48\hat{\mu}%
_{3}\right)  \\
&&-\frac{96}{73}\hat{\mu}_{4}\left( 1491-10800\hat{\mu}_{2}+28864\hat{\mu}%
_{2}^{2}-6240\hat{\mu}_{3}+10752\hat{\mu}_{2}\hat{\mu}_{3}-25088\hat{\mu}%
_{2}^{3}\right) .
\end{eqnarray*}
The last term is the effect of the quartic quasitopological term on $\eta/s$. Clearly it can be either positive or negative;
the investigation of the allowed values of this term  will be given elsewhere.

\section{Concluding Remarks}

We have explicitly constructed the Lagrangian for quartic quasitopological
gravity (up to a term proportional to the Lovelock term) for all dimensions $%
D\geq 5$ except for $D=8$, and shown specifically what its black hole
solutions are. This is the highest-degree case for which it is possible to
find explicit solutions.

It is possible to make some general remarks about quasitopological gravity
even though the specific Lagrangian has not been found for an arbitrary
power $K$ of the curvature. Since all derivative terms higher than 2 must be
eliminated from the Lagrangian, which itself must be linear in the lapse
function, it is reasonable to conjecture that the action in the spherically
symmetric case will be reduce to
\begin{equation}
I_G = \int dt dr N(r) \left(r^{n} \sum_{k=0}^K \hat{\mu}_k \psi^k
\right)^\prime  \label{actpsi}
\end{equation}
up to terms proportional to the transverse volume $V_{n-1}$, for $K$-th
order quasitopological gravity in $(n+1)$ dimensions, where $%
\psi=l^{2}r^{-2}(k-f)$ and the $\hat{\mu}_k$ parameters are rescaled
coefficients of the $k$-th powered curvature term. For a given $K$ this
action should be valid for all dimensionalities $(n+1)$ larger than 4,
except for particular choices where $n=2K-1$.

A similar conjecture was formalized by considering the invariant \cite
{Oliva:2011xu,Oliva:2010zd}
\begin{equation}
\mathcal{N}^{(K)} = \delta^{\mu_1\nu_1\cdots
\mu_K\nu_K}_{\alpha_1\beta_1\cdots \alpha_K\beta_K} \left(R_{\mu_1\nu_1}^{%
\phantom{\mu_1\nu_1}\alpha_1\beta_1} \cdots R_{\mu_K\nu_K}^{%
\phantom{\mu_K\nu_1}\alpha_K\beta_K} - C_{\mu_1\nu_1}^{\phantom{\mu_1\nu_1}%
\alpha_1\beta_1} \cdots C_{\mu_K\nu_K}^{\phantom{\mu_K\nu_1}\alpha_K\beta_K}
\right)
\end{equation}
where $\delta^{\mu_1\nu_1\cdots \mu_K\nu_K}_{\alpha_1\beta_1\cdots
\alpha_K\beta_K}$ is the generalized Kronecker-delta tensor and $C_{\mu\nu}^{%
\phantom{\mu\nu}\alpha\beta}$. When rewritten in terms of Riemann
invariants, $\mathcal{N}^{(K)}$ factorizes with a common factor of $(n-2K+2)$%
; for $n<2K-2$ it vanishes. Taking the action to be of the form
\begin{equation}
I = \int d^{n+1}x \sqrt{-g} \left[ \alpha^{(K)}_0 \frac{n-1}{2^K (n-2K+2)}
\mathcal{N}^{(K)} +\sum_{j=1}^{N^{(K)}_n} \alpha^{(K)}_j W^{(K)}_j \right]
\label{ORact}
\end{equation}
where $\{ W^{(K)}_1 ,\ldots , W^{(K)}_{N^{(K)}} \}$ is a set of linearly
independent $K$-th order Weyl invariants, it is possible to prove a
generalized Birkhoff's theorem provided $\alpha^{(K)}_0$ is an appropriately
chosen linear combination of the $\alpha^{(K)}_j $ coefficients \cite
{Oliva:2011xu}. All the contractions of $K$ Weyl tensors are proportional on
spherical/planar/hyperbolic symmetric spacetimes, with no static assumption
required.

For $n \geq 2K-1$ the invariant $\mathcal{N}^{(K)}$ can be expressed as a
linear combination of the $2K$-dimensional Euler density and all conformal
invariants. In general this action will yield field equations greater than
second order, but for the aforementioned linear combination of the $%
\alpha^{(K)}_j $ coefficients they reduce to second order in the
spherical/planar/hyperbolic cases \cite{Oliva:2011xu}. For $n < 2K-2$ the
action (\ref{ORact}) yields a set of fourth order field equations for an
arbitrary metric but a set of vanishing field equations on
spherical/planar/hyperbolic symmetric spacetimes \cite{Oliva:2011xu}.

In view of our results for the quartic case, we propose that there exists $K$%
-th order quasitopological gravity in any dimension except for $n=2K-1$. For
the spherical/planar/hyperbolic ansatz (\ref{metric}) we conjecture that the
nonvanishing action is given in (\ref{actpsi}), which yields the field
equations
\begin{equation}
\left(r^{n} \sum_{k=0}^K \hat{\mu}_k \psi^k \right)^\prime = 0
\label{genpsi}
\end{equation}
for the metric function $f(r)$. In general the field equations will be of
4th order, since the variation of the action will produce terms proportional
to 2nd derivatives of variations of metric functions multiplied by powers of
the Riemann curvature. Upon integration by parts the largest number of
derivatives that could act on any term will be 4.

This equation has the same form as the corresponding situation in Lovelock
gravity \cite{Camanho:2011rj}, the difference being that $K \leq \left[
\frac{n}{2} \right] $ in the Lovelock case, whereas $K$ is not restricted in
the quasitopological case. The solutions to (\ref{genpsi}) are given by the
solutions to the equation
\begin{equation}
\sum_{k=0}^K \mu_k \psi^k = \frac{m \ell^n}{r^n}  \label{gensol}
\end{equation}
which for $K \geq 5$ cannot be written explicitly in general. The analysis
of the black hole solutions for this case completely parallels that of the
Lovelock case \cite{Camanho:2011rj} and we shall not repeat it here.

While the quartic Lagrangian (\ref{X3}) we have constructed is unique (up to a term
proportional to the Euler density) insofar as it yields second order differential equations for
spherically symmetric metrics,  its geometrical origins remain somewhat obscure. Since
all spherically symmetric metrics reduce to effective theories of gravitation in two space-time dimensions,
it may be that some kind of theorem of principle will single out the choice (\ref{X3}) with coefficients given
in the Appendix.  This remains an interesting topic for future study.

Quasitopological gravity provides a much broader range of parameter space
for holographic duality. It would be interesting to see what constraints are
placed on the entropy-to-viscosity ratio for this class of theories, and how
they modify their condensed matter duals in asymptotically Lifshitz gravity.

\section*{Acknowledgements}

This work was supported by the Research Institute for Astrophysics and
Astronomy of Maragha and the Natural Sciences and Engineering Research
Council of Canada.

\section{Appendix}

Here we present the coefficients of quartic-curvature terms in Eq. (\ref{X4}%
) in $n+1$ dimensions. Using spherically symmetric metric (\ref{metric}),
one can show that the Lagrangian (\ref{X4}) with\ the following $c_{i}$'s
\begin{eqnarray*}
c_{1} &=&-\left( n-1\right) \left( {n}^{7}-3\,{n}^{6}-29\,{n}^{5}+170\,{n}%
^{4}-349\,{n}^{3}+348\,{n}^{2}-180\,n+36\right) , \\
c_{2} &=&-4\,\left( n-3\right) \left( 2\,{n}^{6}-20\,{n}^{5}+65\,{n}^{4}-81\,%
{n}^{3}+13\,{n}^{2}+45\,n-18\right) , \\
c_{3} &=&-64\,\left( n-1\right) \left( 3\,{n}^{2}-8\,n+3\right) \left( {n}%
^{2}-3\,n+3\right) , \\
c_{4} &=&-{(n}^{8}-6\,{n}^{7}+12\,{n}^{6}-22\,{n}^{5}+114\,{n}^{4}-345\,{n}%
^{3}+468\,{n}^{2}-270\,n+54), \\
c_{5} &=&16\,\left( n-1\right) \left( 10\,{n}^{4}-51\,{n}^{3}+93\,{n}%
^{2}-72\,n+18\right) , \\
c_{6} &=&--32\,\left( n-1\right) ^{2}\left( n-3\right) ^{2}\left( 3\,{n}%
^{2}-8\,n+3\right) , \\
c_{7} &=&64\,\left( n-2\right) \left( n-1\right) ^{2}\left( 4\,{n}^{3}-18\,{n%
}^{2}+27\,n-9\right) , \\
c_{8} &=&-96\,\left( n-1\right) \left( n-2\right) \left( 2\,{n}^{4}-7\,{n}%
^{3}+4\,{n}^{2}+6\,n-3\right) , \\
c_{9} &=&16\left( n-1\right) ^{3}\left( 2\,{n}^{4}-26\,{n}^{3}+93\,{n}%
^{2}-117\,n+36\right) , \\
c_{10} &=&{n}^{5}-31\,{n}^{4}+168\,{n}^{3}-360\,{n}^{2}+330\,n-90, \\
c_{11} &=&2\,(6\,{n}^{6}-67\,{n}^{5}+311\,{n}^{4}-742\,{n}^{3}+936\,{n}%
^{2}-576\,n+126), \\
c_{12} &=&8\,{(}7\,{n}^{5}-47\,{n}^{4}+121\,{n}^{3}-141\,{n}^{2}+63\,n-9), \\
c_{13} &=&16\,n\left( n-1\right) \left( n-2\right) \left( n-3\right) \left(
3\,{n}^{2}-8\,n+3\right) , \\
c_{14} &=&8\,\left( n-1\right) \left( {n}^{7}-4\,{n}^{6}-15\,{n}^{5}+122\,{n}%
^{4}-287\,{n}^{3}+297\,{n}^{2}-126\,n+18\right) ,
\end{eqnarray*}
reduces to the Lagrangian given in Eq. (\ref{Igfin2}).

\end{document}